\documentstyle[12pt,fleqn,aasms4]{article}

\newcommand{\al}{\mbox{$\alpha$}}
\newcommand{\p}{{\displaystyle \partial}}
\newcommand{\lnl}{{\displaystyle \ln{L}}}
\newcommand{\rbra}{\mbox{$\left< u_{r}^{2} \right>$}}
\newcommand{\tbra}{\mbox{$\left< u_{t}^{2} \right>$}}
\newcommand{\bra}{\mbox{$\left< u_{i}^{2} \right>$}}
\newcommand{\fbra}{\mbox{$\left< u_{i}^{4} \right>$}}

\newcommand{\lra}[2]{\mbox{$ (#1 \leftrightarrow #2) $}}
\newcommand{\pardif}[2]{{\displaystyle \frac{{\partial}^{2}\ln{L}}{\partial #1 \partial #2}}}
\newcommand{\pardiff}[2]{{\displaystyle \frac{{\partial}^{2}\ln{{\cal L}}}{\partial #1 \partial #2}}}
\newcommand{\vev}[1]{\mbox{$\left< #1 \right>$}}
\newcommand{\be}{\begin{equation}}
\newcommand{\ee}{\end{equation}}
\newcommand{\bea}{\begin{eqnarray}}
\newcommand{\eea}{\end{eqnarray}}
\newcommand{\lab}[1]{\label{#1}}
\newcommand{\r}[1]{~(\ref{#1})}
\newcommand{\kms}{{\rm km}\,{\rm s}^{-1}}

\begin{document}

\title{Mathematics of Statistical Parallax and the Local Distance Scale}
\author{Piotr Popowski, Andrew Gould\altaffilmark{1}\altaffiltext{1}{Alfred P.\ Sloan Foundation Fellow}}
\affil{Ohio State University, Department of Astronomy, Columbus, OH 43210}
\affil{E-mail: popowski,gould@astronomy.ohio-state.edu}

\begin{abstract}
        We present a mathematical analysis of the statistical parallax 
method. The method yields physical insight into the maximum-likelihood 
determinations of the luminosity and velocity distribution and 
enables us to conduct a vigorous Monte Carlo investigation into various 
systematic effects. We apply our analytic formalism to the RR Lyrae sample 
of Layden et al. 
The velocity distribution of 
RR Lyrae stars is highly non-Gaussian, 
with kurtoses $K_{\pi}=2.04$, $K_\theta=3.22$ and $K_z=4.28$ in the three 
principal directions, but this has almost no effect on either the best fit or 
the uncertainty of the luminosity determination. Indeed, our principal
result is that the statistical parallax method is extremely robust in the 
face of all systematic effects that we considered.

Our analysis, applied to the Layden et al.\ RR Lyrae sample, strictly confirms 
the majority of their results.
The mean RR Lyrae absolute magnitude is $M_V=0.75\pm 0.13$ at the
mean metallicity of the sample $\left<\rm [Fe/H]\right>=-1.61$, compared to 
$M_V=0.71\pm 0.12$ obtained by Layden et al.  Most of the difference is due 
to Malmquist bias which was not considered in previous studies.
We also analyze a semi-independent non-kinematically selected sample of stars 
with metallicities at the
[Fe/H]$\leq -1.5$ taken from Layden et al.\ and Beers \& Sommer-Larsen 
and obtain $M_V=0.79\pm 0.12$ at $\left<\rm [Fe/H]\right>=-1.79$.
Additionally, this analysis yields measurements of the radial bulk
motion ($4\pm 10 \,\kms$) and 
vertical bulk motion ($0\pm 6 \,\kms$) of the halo relative to the 
Local Standard of Rest.

\end{abstract}

\section{Introduction}
The discrepancy between the Cepheid and RR Lyrae distance estimates to the 
Large Magellanic Cloud (LMC) is a long-standing problem (e.g., van den Bergh 
1995).
There are two possible sources of disagreement between the Cepheid and RR Lyrae
distance scales: either the compared populations of Cepheids or RR Lyrae 
stars in the Milky Way and LMC are different or there are substantial errors 
in the Cepheid or RR Lyrae calibration. 
In this paper we investigate the statistical and systematic errors associated 
with the statistical parallax method, which is one of the main calibration
methods for RR Lyrae stars.
We prove that the statistical parallax method is very robust.
Consequently, miscalibration of the absolute magnitude of RR Lyrae stars is not
likely to be the solution to the distance scale problem and it is fair to state
that RR Lyrae stars constitute reliable distance indicators.
We conclude that either the Cepheid distance scale is incorrect or the
currently available data do not allow for a reliable comparison between
RR Lyrae stars in the Milky Way and LMC.

        Historically, the methods of secular parallax and 
classical statistical parallax (e.g., Trumpler \& Weaver 1962) 
have been used to expand the method of
trigonometric parallax to more distant objects by increasing the baseline
of the measurement.  Secular parallax takes advantage of the fact that the 
Sun moves at a speed $W\sim 20\,\kms$ relative to disk populations and 
$W\sim 200\,\kms$ relative to halo populations, corresponding to 4 AU per year
and 40 AU per year, respectively.  After a time $t$ of several 
decades, the accumulated baseline $W t$  is several orders of magnitude larger
than the Earth-Sun baseline used in trigonometric parallax.  
If the speed of
the {\it individual star} (rather than of the population
as a whole) were known to be $W$, then one could measure the distance with a
precision better by a factor $W t/\rm AU$ than using trigonometric parallax 
for the same astrometric accuracy.  The drawback is that the stars have
a dispersion $\sigma$, so that even in the limit of perfect measurements,
the precision of the distance determination is inversely proportional to the
``Mach number'', $\kappa=W/\sigma$, which is typically of order 1.  
By applying this method to a sample of $N$ stars, one can beat down the
noise and achieve a precision $\propto N^{-1/2}\kappa^{-1}$.  The 
velocity dispersion itself can be used to obtain an independent measurement 
of the distance using the method of classical statistical parallax.  That is, 
the distance scale of a class of stars can be estimated by forcing their 
radial velocities and proper motions to reproduce the same velocity 
dispersion. The precision of this method is $\propto N^{-1/2}$. The
two determinations can be combined to achieve a joint precision
$\propto [N(1+ g\kappa^2)]^{-1/2}$, where $g\simeq 1/6$ is a geometrical 
factor that we derive below.  Nevertheless, because the method is limited
by the number of stars rather than the precision of the measurements, its
usefulness has been restricted to ever more distant classes of stars as
astrometry has improved.

        Only recently (e.g., Murray 1983; Hawley et al.\ 1986; Strugnell,
Reid, \& Murray 1986) was it fully realized that secular parallax and
classical statistical parallax are actually two aspects of the same
generalized method, now simply called ``statistical parallax''.  
Secular parallax is based on forcing equality 
between the three first moments of the velocity distribution 
(the bulk motion $\bf W$) as determined from 
radial velocity and proper motion measurements, while classical statistical
parallax is based on forcing equality of the six second moments
(the six independent components of the velocity covariance matrix $C_{i j}$).  
In the modern combined version of statistical parallax one simply determines 
ten parameters simultaneously by applying maximum likelihood.  The ten
parameters are an overall distance scaling factor $\eta$ (relative to an 
initial arbitrary distance scale) plus the nine first and second moments,
$\bf W$ and $C_{i j}$.

Hawley et al. (1986) and Strugnell, Reid, \& Murray (1986) were
the first ones to use a maximum likelihood analysis to determine these 10
parameters. Actually both studies use 11 parameter models of the form
introduced by Murray (1983), but hold
fixed the 11th parameter, the dispersion in $\eta$. 
As we show in \S 3.1, current data do not allow one to put any useful 
constraint on this parameter, and we remove it from our primary analysis.
The dispersion in $\eta$ also gives rise to Malmquist (1920) bias. However,
in the analysis conducted by Hawley et al. (1986) and Strugnell et al. (1986),
the inclusion of the dispersion in $\eta$ does not by itself
correct for Malmquist bias. Correcting for the Malmquist bias requires 
additional modeling of the selection criteria of the stars in the sample and
of the distribution of absolute magnitudes. It is most effectively done 
on a star by star basis (for a discussion see e.g. Smith 1987, also
Ratnatunga \& Upgren 1997). 
However when the detailed selection criteria are not available and even the 
magnitude of the scatter is uncertain, such procedures are no better than 
correcting the final result for the ``average'' Malmquist bias 
(e.g. Jung 1970).
\footnote{Actually, for small scatter in absolute magnitude, the two
approaches are equivalent because they differ only by negligible second
order terms.}

Ideally, the statistical parallax method should be applied to a group of
stars that are:
\begin{enumerate}
\item dynamically homogeneous, i.e. all the stars are drawn from a single
velocity distribution (not constrained to be Gaussian) regardless of their
locations
\item standard candles, i.e. all have the same absolute magnitude.
\end{enumerate}
Theoretically, condition (1) can be met by careful selection of stars
in the nearby solar neighborhood. In practice, to obtain a statistically 
satisfactory sample often requires that stars be collected from a region 
comparable ($\sim 3-5$ smaller) in size to the distance to Galactic center. 
In such a case one does not expect condition (1) to be exactly satisfied. 
However, one can reasonably assume that in this part of the Galaxy the bulk 
motion and velocity ellipsoid change monotonically with distance from
Galactic center. Hence, it is important to probe parts of the sky 
towards and away from the Galactic center. 
Although it is difficult to check whether condition (1) is satisfied, 
it is possible to investigate some of the possible systematic errors 
analytically and to quantify others through Monte Carlo simulations.
Condition (2) must be relaxed somewhat, and it suffices to have
stars with a small scatter around calibrated absolute magnitudes, as we 
discussed above and as we further discuss in \S 3.

The RR Lyrae stars of the Galactic spheroid constitute a very good
sample for analysis by statistical parallax.
They are close to satisfying  
condition (1), being abundant in the solar vicinity and distributed 
reasonably evenly in the sky.
Their absolute magnitudes can be obtained from the
absolute magnitude -- metallicity relation (e.g., Carney et al. 1992) with 
a scatter not likely to exceed 0.15 magnitudes (condition 2).
Additionally, RR Lyrae stars can be observed not only in the Milky Way, but 
also in neighboring galaxies like the LMC.

The organization of this paper is as follows.
In \S 2 we analyze a simple model of a stellar system with an isotropic 
velocity ellipsoid and we estimate the error in the distance scaling parameter 
$\eta$.
We also discuss the role of observational errors and the systematic error
caused by the mis-estimation of these observational errors.
In \S 3, we apply the results of \S 2 to investigate various possible 
systematic effects that may bias the statistical parallax solution. 
In \S 4, we analyze the full maximum likelihood formulation of the problem
for the limiting case of negligible measurement errors.
This case mimics the actual situation because the intrinsic velocity
dispersion, being much larger than the observational errors, dominates
the statistical uncertainty.
We obtain algebraic expressions for the uncertainty in all 10 parameters, 
$\eta$, $W_{i}$, and $C_{ij}$.
In particular, the error in $\eta$ for the general case has the same form
as it does for the naive model of \S 2. That is, as we anticipate in
\S 2, the seemingly complicated ``black box'' of maximum likelihood can 
be understood in simple physical terms. 
In \S 5 we conduct the complete maximum likelihood analysis of the 
general case including observational errors in stellar velocities.
We then reanalyze the Layden et al. (1996) sample of 162 halo stars found in 
the solar neighborhood and confirm most of their findings.
In \S 6 we conduct Monte Carlo simulations to confirm our analytic 
results and find possible biases, e.g., those induced by the anisotropic
spatial distribution of the stars in the sample.
In \S 7 we analyze a sample of halo stars constructed by combining 
the samples of Layden et al. (1996) and Beers \& Sommer-Larsen (1995).
Finally, in \S 8 we summarize our results and discuss their implications for 
the local distance scale.
In particular, we conclude that the statistical parallax method is very robust
against various systematic effects.

\section{Isotropic velocity dispersion}

In this section we conduct the heuristic analysis of the statistical
parallax method. It is aimed at explaining the physical foundations
of this powerful method and gives the basis for the analysis of possible 
systematic errors. 
In \S 4 and \S 5, we develop a more rigorous mathematical treatment 
which is then used in our Monte Carlo simulations described in \S 6.
As mentioned in \S 1, in the general case there are 10 parameters of the fit: 
the distance scaling parameter $\eta$, 3 components of bulk motion $W_{i}$, 
and 6 independent components of symmetric velocity covariance matrix $C_{ij}$.
We initially consider a stellar system with an isotropic velocity ellipsoid.
The number of independent components of $C_{ij}$ is thereby reduced
from 6 to 1: $C_{ij}={\sigma}^{2} {\delta}_{ij}$.

We define the scaling parameter $\eta$ by
\be
\eta={\left( \frac{L_{true}}{L_{assumed}} \right)}^{\frac{1}{2}}, \lab{eta.def}
\ee
where $L_{true}$ is the actual luminosity of the chosen class of star 
(e.g., RR Lyrae), and $L_{assumed}$
is its assumed luminosity, which can be chosen arbitrarily for purposes
of making the calculation.

Below we estimate analytically the error associated with the determination
of the distance scaling parameter $\eta$. 
When dealing with errors we shall apply the general rule that combining 
errors of statistically independent quantities we add their variances to 
get the total variance, computing the resultant error of two statistically
independent measurements of the same quantity, we add variances harmonically.

We decompose the velocity of each star into its radial and tangential
components, ${\bf u}=(u_{r},{\bf u}_{t})$.
Now we can conveniently separate two sources of information about $\eta$.
The first comes from forcing equality between the velocity dispersions as 
determined from radial and transverse velocities (``classical statistical
parallax''). The second comes from forcing equality in the bulk motion 
inferred from radial and transverse velocities (``secular parallax'').
We assume that these two estimates of $\eta$ are independent in the
statistical sense.
To simplify the analysis we initially assume that
there is no bulk motion, which implies that $\vev{{\bf u}}=0$, where the 
brackets, here and afterwards, symbolize averaging unless otherwise noted.
Thus 
\be
\tbra=2 \rbra, \lab{vt.square}
\ee
where $u_{t}=|{\bf u}_{t}|$.
The inferred transverse speed ${\tilde{u}}_{t}$ is related to the proper 
motion $\mu$ of a star by:
\be
{\tilde{u}}_{t}=\mu \cdot d_{assumed}= \mu \cdot \frac{d_{true}}{\eta}= \frac{u_{t}}{\eta}, \lab{vt.tilde1}
\ee
where $d_{true}$ and $d_{assumed}$ are the true distance and the distance
inferred based on the assumed luminosity, respectively.

We combine equations\r{vt.square} and\r{vt.tilde1} to express $\eta$ as
\be
{\eta}^{2}=2 \frac{\rbra}{\vev{{\tilde{u}_t}^{2}}} \lab{eta.square}
\ee

Each component of the velocity, $u_{i}$, of a given star may be 
regarded as being drawn from a one-dimensional distribution $f(u_{i})$. 
The fractional error in the velocity dispersion based on a sample
of $N$ stars is
\be
\frac{\Delta(u_{i}^{2})}{\bra} \equiv \frac{{\left[ var \left( u_{i}^{2} \right)
\right]}^{\frac{1}{2}}}{\bra}={\left( \frac{K-1}{N} \right)}^ {\frac{1}{2}}, 
\lab{disp1}
\ee
where in the last step we have used the definition of the kurtosis $K$ for
$\vev{u_{i}}=0$:
\be
K=\frac{\fbra}{{\bra}^2}
\ee
By equation\r{disp1}
\be
\frac{\Delta(u_{r}^{2})}{\rbra}={\left( \frac{K-1}{N} \right)}^ {\frac{1}{2}}.
\lab{error.v}
\ee
Recall that ${\tilde{u}_t}^{2}$ in equation\r{eta.square} designates the 
sum of the squares of the transverse velocities in two perpendicular 
directions. Thus
\be
\frac{\Delta({\tilde{u}_t}^{2})}{\vev{{\tilde{u}_t}^{2}}}={\left( \frac{K-1}{2N} \right)}^ {\frac{1}{2}}. \lab{error.t}
\ee
Adding the fractional errors from formulae\r{error.v} and\r{error.t} in 
quadrature gives the fractional error in $\eta$ (see eq.\r{eta.square}),
\be
{\left. \frac{\Delta \eta}{\eta} \right|}_{disp}={\left( \frac{3(K-1)}{8N} \right)}^ {\frac{1}{2}} \;\;\; 
\stackrel{\mbox{{\tiny\rm \raisebox{0.1cm}{Gaussian}}}}{\mbox{{\Huge \mbox{$\longrightarrow$}}}}
 \;\;\; {\left( \frac{3}{4N} \right)}^ {\frac{1}{2}} \lab{eta.error1}
\ee
where we have used $\Delta \eta / \eta = (1/2) \Delta({\eta}^{2}) / {\eta}^{2}$, and where in the last step we have evaluated the expression for the case
of a Gaussian distribution.

The bulk motion of the whole sample relative to the Sun gives independent
information about $\eta$. For simplicity we assumed in the above 
analysis that the bulk motion was zero, but now we relax this assumption.
In analogy to equation\r{eta.square} we can write
\be
\eta=\frac{W_{i,r}}{W_{i,t}}, \lab{eta2}
\ee
where the index $i$ indicates the component and the indices $r$ and $t$ mean 
``as inferred from radial and transverse velocities'', respectively.
%
In general, we decompose the bulk motion into components in an arbitrary frame
of reference which generally results in all three components having non-zero 
values. Here for the sake of simplicity, we pick the frame of reference with
z-axis aligned with the bulk motion of the sample.
We restrict attention to the radial velocities of the stars but the results 
are representative of each of the components.
One can express the radial velocity as the bulk motion along the line
of sight with uncertainty equal to the velocity dispersion $\sigma$:
\be
u_{r}=W_{3,r}\cos{\theta}\pm\sigma \lab{v}
\ee
where $W_{3,r}$ represents the bulk motion component
inferred from the radial velocity.
We may construct an estimator of any of the bulk motion components 
(in this simplified case of $W_{3,r}$) by dividing equation\r{v} (or its
generalized form) by the 
appropriate angular dependence:
\be
W_{3,r}=\frac{u_{r}}{\cos{\theta}} \mp \frac{\sigma}{\cos{\theta}} 
\lab{w3r.est}
\ee
For the sample of $N$ stars randomly distributed in the sky, the fractional
error in estimating $W_{3,r}$ is
\be
\frac{\Delta W_{3,r}}{W_{3,r}} = {\left[ {\sum}_{i=1}^{N} \frac{{\cos}^{2} \theta}{{\sigma}^{2}} \right]}^{-\frac{1}{2}} \longrightarrow {\left( \frac{3}{N}
\right)}^{\frac{1}{2}} \frac{\sigma}{W_{3,r}}, \lab{var.w3r}
\ee
where in the last step we have used the fact that
$\vev{{\cos}^{2} \theta}=1/3$ averaged over all the angles.
The transverse velocities contain twice as much information as the radial ones,
so
\be
\frac{\Delta W_{3,t}}{W_{3,t}} =
{\left( \frac{3}{2N}\right)}^{\frac{1}{2}} \frac{\sigma}{W_{3,t}}.
\lab{var.w3t}
\ee
Applying equation\r{eta2} and adding the errors from equations\r{var.w3r} and\r{var.w3t} in quadrature, we find that
\be
\frac{\Delta \eta}{\eta}={\left( \frac{9}{2N} \right)}^{\frac{1}{2}} \frac{\sigma}{W_{3}}.
\ee
A similar analysis can be made to extract information from $W_{1}$ and $W_{2}$.
Since these three pieces of information about $\eta$ are independent, 
we add variances harmonically and obtain
\be
{\left. \frac{\Delta \eta}{\eta} \right|}_{bulk} = {\left( \frac{9}{2N} \right)}^{\frac{1}{2}} {\kappa}^{-1}, \;\;\;\;\;\; \kappa\equiv\frac{W}{\sigma}.
\lab{eta.error3}
\ee
where $\kappa$ is the ``Mach number'', the ratio of bulk motion to dispersion.
Combining equations\r{eta.error1} and\r{eta.error3} yields
\be
\frac{\Delta \eta}{\eta}= N^{-\frac{1}{2}} {\left( \frac{8}{3(K-1)}+\frac{2}{9} {\kappa}^2 \right)}^{-\frac{1}{2}} \;\;\;
\stackrel{\mbox{{\tiny\rm \raisebox{0.1cm}{Gaussian}}}}{\mbox{{\Huge \mbox{$\longrightarrow$}}}} \;\;\;
N^{-\frac{1}{2}} {\left( \frac{4}{3}+\frac{2}{9} {\kappa}^2 \right)}^{-\frac{1}{2}}. 
\lab{eta.error4}
\ee
The typical values for population II stars in the solar 
vicinity are $W \sim 200\, \rm{km} \, {\rm s}^{-1}$ and $\sigma \sim 
100\, \rm{km} \, {\rm s}^{-1}$, $\kappa \sim 2$ so $(2/9){\kappa}^2 \sim 8/9$. 
Hence, 60\% of the information about the distance scale 
comes from velocity dispersions and 40\% from the bulk motion.

\subsection{Measurement Errors}

	Here we investigate how measurement errors affect the statistical 
accuracy.  We adopt the simplified model discussed above with isotropic 
dispersion $\sigma$, bulk motion $W$, typical measurement errors ${\sigma}_r$ 
and $\sigma_\mu$, and typical stellar distances $D$.  
We define $\delta_r\equiv {\sigma}_r/\sigma$ and 
$\delta_\mu\equiv D\sigma_\mu/\sigma$.
If $\eta$ is determined solely from the bulk motion, then the accuracy is
${(\Delta \eta/\eta)}^{2} = 3 N^{-1}{\kappa}^{-2}
[(1+\delta_r^2)+0.5(1+\delta_\mu^2)]$.  On the other hand if $\eta$
is determined from the dispersion,
${(\Delta \eta/\eta)}^{2} = 0.5 N^{-1}([(1+\delta_r^2)+0.5(1+\delta_\mu^2)]$.  Thus,
the accuracy of the overall measurement is degraded by a fractional amount,
\be
{\delta (\Delta \eta/\eta)\over (\Delta \eta/\eta)}\sim {\delta_r^2\over 3} + 
{\delta_\mu^2\over 6}\sim 0.04,
\ee
where we have adopted $\delta_r\sim 0.2$ and $\delta_\mu\sim 0.4$ appropriate
for RR Lyrae stars.  Hence, realistic measurement errors have an extremely
small effect on the precision of the luminosity determination, a result which
we confirm numerically in \S 6.  This justifies
our approach of ignoring measurement errors in our analytic investigation
of various other effects conducted in \S 3.

Most generally, the uncertainty in the statistical parallax determination is 
set by:
\begin{enumerate}
\item the intrinsic velocity dispersion of the system
\item the bulk motion of the sample
\item the observational errors
\item the size of the sample
\end{enumerate}
Once the stellar system is given we have no control over items 1 \& 2, but
we can still vary items 3 \& 4 to minimize the uncertainty.
In the specific case of negligible observational errors (meaning small relative
to intrinsic velocity dispersion as in the considered case), the only 
remaining factor influencing the uncertainty is the size of the sample.
That is, the only way to substantially reduce the uncertainty
is to increase the sample size and not decrease the observational errors.

\subsection{Mis-estimate of the Observational Errors}

	As shown in \S 2.1, observational errors increase only slightly
the uncertainty in the estimate of $\eta$.  Observational errors can also
result in a biased estimate of $\eta$ if they are not properly accounted
for. Suppose that the radial velocity errors
are ${\sigma}_r$, but that they are taken to be zero in the analysis.
(For simplicity, we initially assume that there are no proper-motion errors.)
If $\eta$ is estimated by comparing the radial and proper-motion
dispersions, the radial dispersion will be overestimated by a factor
$1+({\sigma}_r/\sigma)^2$ while the proper-motion dispersion will be properly
estimated.  Hence $\eta$ will be overestimated by a fraction
$(\delta\eta/\eta)=[1+({\sigma}_r/\sigma)^2]^{1/2}-1\sim ({\sigma}_r/\sigma)^2/2$.  
On the other hand, there will be no effect on the estimate of $\eta$ based on 
comparing the radial-velocity and proper-motion bulk motion.  Hence the total 
systematic error in the combined determination will be the weighted average 
of the two:
$(\delta\eta/\eta)\sim ({\sigma}_r/\sigma)^2/[2+{\kappa}^2/3]\sim 
0.3({\sigma}_r/\sigma)^2$.  If the true observational error is ${\sigma}_r$, but 
the analysis incorrectly treats the error as $\tilde{\sigma}_r$, then
$(\delta\eta/\eta)\sim 0.3\delta{\sigma}_r^2/\sigma^2$, where
$\delta{\sigma}_r^2\equiv{\sigma}_r^2-\tilde{\sigma}_r^2$.  Finally, we allow for
a mis-estimate of the proper-motion errors and define
$\delta\sigma_\mu^2\equiv\sigma_\mu^2-\tilde\sigma_\mu^2$ in analogy to the
radial-velocity term.  We take the typical distance of stars in the sample
to be $D$ and estimate a net systematic error of
\be
\frac{\delta\eta}{\eta} \sim 0.3\frac{\delta{\sigma}_r^2-D^2\delta\sigma_\mu^2}{\sigma^2}.
\ee
Since ${\sigma}_r/\sigma\sim 0.2$ and $D\sigma_\mu/\sigma\sim 0.4$, the most
likely source of a major effect is mis-estimation 
of the proper motion errors,
but even this is not likely to be large.  For most RR Lyrae stars in the sample,
the proper-motion error is estimated to be 5 mas yr$^{-1}$.  Suppose that
the true value is 20\% higher.  Then $\eta$ would have 
been underestimated by only $(\delta\eta/\eta) \sim 1.7\%$.

\section{Analytic Investigation of Systematic Effects}

\subsection{Dispersion of the Luminosity}

        Throughout our treatment, we have assumed and will assume
in \S 4 and \S 5 that the entire population
of stars has exactly the same luminosity.  This is not customary.  One
can in principle fit for the dispersion in luminosity (in which case there
are 11 fit parameters instead of 10) or one can assume a certain dispersion
which then affects the values of the remaining 10 parameters (e.g.\ Hawley
et al.\ 1986; Strugnell et al.\ 1986; Layden et al.\ 1996).  
Here we show that for the RR Lyrae sample, there are several orders of 
magnitude too little information to determine the dispersion from the data as 
was also suggested by previous studies.  Moreover, we show that the effect 
of including the dispersion is one order of magnitude smaller than 
the statistical errors, and that one is therefore justified in accounting for 
this effect separately.

        We assume that the stars have a range of luminosities which we 
parameterize as a dispersion in absolute magnitude $M_V$, $\sigma_M$, 
but that the analysis
is conducted assuming that all stars have the same luminosity.  We work within
the simplified framework of \S 2.  The inferred distance to each star in
the best-fit solution will then deviate from the true distance by 
${\cal O}(\sigma_M)$. This will increase both the variance in the inferred
transverse speeds and the mean.  Averaged over all 
directions, the fractional increase of the dispersion is 
$2(\ln 10/5)^2[1 + {\kappa}^2/6]\sigma_M^2$.  
Thus, if $\eta$ were determined by matching the dispersions of the radial
velocities and proper motions, it would be underestimated by a fraction
$(\delta\eta/\eta) = -(\ln 10/5)^2[1 + {\kappa}^2/6]\sigma_M^2
\sim -0.4\sigma_M^2$.  
On the other hand, if $\eta$ were determined by matching the bulk motions,
it would be underestimated by a fraction 
$(\delta\eta/\eta) = - 0.5(\ln 10/5)^2\sigma_M^2 \sim -0.1\sigma_M^2$.  
Thus, by comparing the two estimates,
one could in principle determine $\sigma_M$.  For the RR Lyrae sample, 
however, the precision of each method is only $\sim 7\%$, so that the
precision of the difference is $\sim 10\%$.  This means that the data set
would have to be increased $\sim 400$ fold in order to detect plausible
values of $\sigma_M\leq 0.15\,$mag.  If the dispersion is simply ignored and
the two methods are averaged (as they automatically are in the maximum 
likelihood approach),
then $\eta$ will be underestimated by
$(\delta\eta/\eta) = -(\ln 10/5)^2[12 + 3{\kappa}^2]\sigma_M^2/[12 + 2{\kappa}^2]
\sim -0.25\sigma_M^2$. 
This is an extremely small correction for plausible values of $\sigma_M$.
The underestimate in $\eta$ in turn induces changes of the bulk
motion and dispersions, which we estimate numerically in \S 6.

\subsection{Malmquist bias}
 
       Malmquist bias is another correction to $\eta$ which scales
$\propto \sigma_M^2$, but it is of opposite sign.
As we discuss in the introduction,
the Malmquist bias correction is {\it not} automatically taken into
account by simply considering the 11-parameter model, with the
11th parameter being
the dispersion in $\eta$. The assumption intrinsic to most such models (and
also to our 10-parameter model) is that the stars in the sample are
representative of the whole population. This assumption is not correct
for samples that have been selected based on stellar magnitudes.
Malmquist bias must therefore be put by hand\footnote{See e.g. Smith (1987)
for alternative approach, but see also the discussion of its limited
application in \S 1.}.
Malmquist (1920) showed that the mean distance of a magnitude-limited sample 
drawn from a homogeneous distribution with a Gaussian scatter in absolute 
magnitude will be higher than that of one without scatter by 
$(\delta\eta/\eta) \sim 3{(\ln{10/5})}^{2} \sigma_M^2 \sim (\ln{10/5}) 1.38\sigma_M^2  \sim +0.64 \sigma_M^2$.

        Malmquist bias is a particular form of 
``selection bias'' which arises whenever a sample deviates systematically
from the underlying population because of the way that it was selected.  
The magnitude of the selection bias depends, in general, on the form of the 
selection.  The
RR Lyrae sample is ``local'' in the sense that it contains stars that are less
than 2 or 3 kpc from the Sun, but does not contain substantially more distant
stars (even though such stars are plentiful).  Some procedure
(whose details are not known to us) had to be applied to select this sample.
In principle, one could distinguish between nearby and distant stars by 
several methods, including apparent magnitude, trigonometric parallax, and
proper motion.  However, while we do not know exactly
how the sample was selected,
we do know that of these types of information only apparent magnitudes were 
available.  Therefore, each star can have been included only because
it satisfied {\it some} magnitude limit, even if this magnitude limit is
different from star to star.  The correction calculated by Malmquist 
(1920) is independent of the magnitude limit, so the same correction applies
to all sub-samples of fixed magnitude limit, even samples with only one
object.  Therefore, it applies to the sample as a whole.  For an inhomogeneous
underlying population, the numerical pre-factor in 
$3{(\ln 10/5)}^{2}\sigma_M^2$ 
may in general differ from 3.  However, we have confirmed numerically that for
the distribution of RR Lyrae stars from which the Layden et al.\ (1996) sample
is drawn, the pre-factor is in fact very close to 3.

\subsection{Velocity-Position Correlations}

	Statistical parallax makes the implicit assumption that the stars
seen in all directions have the same velocity distribution.  Since the 
RR Lyrae stars in the sample have typical distances $\sim 2\,$kpc, which is a
significant fraction of $R_0\sim 8\,$kpc, it is possible that this assumption
is not valid.  Suppose, for example, that the velocity dispersion in the $z$
direction falls with distance from the Galactic plane.   If one attempted to
fix the luminosity by matching the $z$-dispersions of the radial velocities
and the proper motions, one would underestimate its value.  This is because
the radial-velocity measurements would be made primarily on stars far from the
plane, while the proper-motion measurements would be made primarily on stars
near the plane.  The high dispersion of the latter would be mistakenly
interpreted as indicating that the stars were closer and hence less luminous
than they actually are.  While this systematic bias would be diluted by
unbiased dispersion measurements in the other two directions and by unbiased
measurements of the bulk motion, it would still affect the final result.
(We focus attention on the $z$ direction because that is the only axis for 
which there is a physical plane of symmetry.  Any correlations along the
other two axes would tend to have opposite signs in the positive and negative
directions and hence would cancel to lowest order.)

	To make a quantitative estimate of the size of this effect, we consider
a simple model where the stars are observed over a uniform sphere of radius
$D$, and have a $z$ velocity dispersion, 
$\sigma_z^2(z) = a(1- f|z|/D)$.  Consider an ensemble of stars at 
a distance $r$ and angle $\theta$ from the north Galactic pole.  These will
enter into the radial-velocity and proper-motion estimates of the $z$ 
dispersion with statistical weights $\cos^2\theta$ and $\sin^2\theta$, 
respectively.  If the correct luminosity of the stars were adopted, the
respective dispersion estimates would be
\be
{\int_0^D d r r^2 \int_{-1}^1 d\cos\theta\,\cos^2\theta
a[1 - f (r/D)|\cos\theta|]
\over\int_0^D d r r^2 \int_{-1}^1 d\cos\theta\,\cos^2\theta}
= a\biggl(1 - {9\over 16}f\biggr)
\ee
and
\be
{\int_0^D d r r^2 \int_{-1}^1 d\cos\theta\,\sin^2\theta
a[1 - f (r/D)|\cos\theta|]
\over\int_0^D d r r^2 \int_{-1}^1 d\cos\theta\,\sin^2\theta}
= a\biggl(1 - {9\over 32}f\biggr).
\ee
\def\lsim{{{}_{{}_<}^{~}\atop {}^{{}^\sim}_{~}}}
Thus, one would be led to make an underestimate $(\delta\eta/\eta)=-(9/64)f$.
Since the effect is diluted by measurements of the dispersion in the other 
two directions and of the bulk motion, the net bias is
$(\delta\eta/\eta) = -(3/64)f/[1 + {\kappa}^2/6]\sim -0.03\,f$.  For plausible
values of $f \lsim D/R_0\sim 0.25$, 
this would be a small but perhaps not completely negligible correction.  
	
We have therefore reanalyzed the sample including an 11th parameter,  
the velocity-dispersion gradient in the $z$ direction.  The best fit
scale length
for this gradient is $7\,$kpc, but since the reduction in $\chi^2$ is 
only $\Delta\chi^2=1.15$, the gradient is not statistically significant.  In
any event, even if the gradient is real it introduces a systematic error of
only $(\delta\eta/\eta)=-0.005$.

\section{Analytic Predictions}

In this section we use the maximum likelihood method to analyze the errors 
intrinsic to the statistical parallax method. We again neglect velocity 
measurement errors. We argued in \S 2.1 that for the halo stars of the Galaxy
the accuracy of the determination of the model parameters scales with the size
of the sample of stars and is barely influenced by observational errors.
Having selected the group of stars (e.g., RR Lyrae stars), we choose an 
orthogonal frame of reference and resolve each star's velocity into components
designated as $u_{i}$. We assume that the velocity distribution is a 
three-dimensional Gaussian, but we do not assume isotropy.
At first Gaussianity of the assumed velocity distribution seems to be a rather
restrictive condition, especially in view of the fact that the actual 
distribution is highly non-Gaussian. However, we argue at the end of
this section (\S 4.1), and confirm numerically in \S 6 that no bias is 
introduced by assuming that the distribution is Gaussian.

The probability of finding a star with three velocity components in the
ranges $(u_{i}, u_{i}+du_{i})$ is given by:
\be
L(u_{i}; \eta, C_{ij}, W_{i}) {\left. d^{3}u \right|}_{assumed} = 
\frac{1}{{(2\pi|C|)}^{\frac{1}{2}}} 
\exp{\left[-\frac{1}{2} {\sum}_{i,j} (u_{i}-W_{i}){(C^{-1})}_{ij}(u_{j}-W_{j})\right]}
d^{3}u,
\lab{prob.fun1}
\ee
where ${\bf W}$ is the bulk motion relative to the Sun, $C$ is the velocity
covariance matrix, and $|C|$ is its determinant.
The volume element in three-dimensional true velocity phase space is $d^{3}u$,
whereas ${\left. d^{3}u \right|}_{assumed}$ is the corresponding element
in ``assumed'' velocity phase space. Since two of the ``assumed'' components
are proportional to ${\eta}^{-1}$,
$ d^{3}u={\eta}^{2} {\left. d^{3}u \right|}_{assumed} $.
Hence the probability density of finding a star with the observed velocity 
components in the ranges $(u_{i}, u_{i}+du_{i})$ is given by:
\be
L(u_{i}; \eta, C_{ij}, W_{i}) = \frac{{\eta}^{2}}{{(2\pi|C|)}^{\frac{1}{2}}} 
\exp{\left[-\frac{1}{2} {\sum}_{i,j} (u_{i}-W_{i}){(C^{-1})}_{ij}(u_{j}-W_{j})\right]},
\lab{prob.fun2}
\ee
The logarithm of the total probability of finding the system in the 
observed state is:
\be
\ln{{\cal L}}={\sum}^{N}_{k=1} \ln{L_{k}}, \lab{ln.L}
\ee
where $L_{k}$ is the probability\r{prob.fun2} associated with
the $k^{\rm th}$ star.
The curvature matrix is given by
\be
B_{\al \beta}= -\pardiff{\al}{\beta}= -{\sum}^{N}_{k=1} \frac{{\partial}^{2}
\ln{L_{k}}}{\partial \al \partial \beta}
\ee
where $\al$ and $\beta$ range over the 10 parameters of the model.
The covariances among the parameters are then given by
\be
{\rm cov}(\al, \beta)={\left( {B}^{-1} \right)}_{\al \beta}. \lab{cov}
\ee
We designate the inverse of the covariance velocity matrix as
\be
Q \equiv C^{-1} \lab{Q.def1}
\ee
which is a special case of the general definition\r{Q.def2}.
We express
\be
u_{i} \equiv u_{r} r_{i} + \eta t_i, \lab{u.def1}
\ee
in terms of its radial and transverse components. Note that $t_i$ is the
projection of $\tilde{\bf u}_{t}$, defined in \S 2, on $i$-th direction. 
We also define $v_{i}$ to be the random part of the velocity:
\be
v_{i} \equiv u_{i} - W_{i}. \lab{u.def2}
\ee

We evaluate the curvature matrix in two steps. First we evaluate the 
contribution of each star to equation\r{ln.L}. Next, we sum over all stars 
under the assumption that they are isotropically distributed on the sky.
We find
\be
-\pardif{C_{mn}}{C_{kl}} = -\frac{1}{2}  Q_{lm} Q_{nk} + (Qv)_{n} Q_{lm} (Qv)_{k} + \lra{k}{l} + \lra{m}{n}
\ee
\be
-\pardif{\eta}{C_{kl}} = -(Qt)_{k} (Qv)_{l} + \lra{k}{l}
\ee
\be
-\pardif{W_{s}}{C_{kl}} = Q_{sk} (Qv)_{l} + \lra{k}{l}
\ee
\be
-\frac{{\partial}^{2} \ln{L}}{\partial {\eta}^{2}} = \frac{2}{{\eta}^{2}}+ 
\vev{t|Q|t}
\ee
\be
-\pardif{W_{s}}{\eta} = - (Qt)_{s}
\ee
\be
-\pardif{W_{l}}{W_{s}} = Q_{sl},
\ee
where the symbol $\lra{k}{l}$ means ``add the terms with $k$ and $l$ 
exchanged, but only if $k \ne l$'', and where we have employed Dirac notation,
i.e.
\be
\vev{X|{\cal O}|Y} \equiv {\sum}_{ij} X_{i} {\cal O}_{ij} Y_{j}. 
\lab{Dirac}
\ee 

Averaging over positions leads to:
\be
-\frac{1}{N} \pardiff{C_{mn}}{C_{kl}} = \frac{1}{2}  Q_{lm} Q_{nk} + \lra{k}{l} + \lra{m}{n} \lab{matrix.first}
\ee
\be
-\frac{1}{N} \pardiff{\eta}{C_{kl}} = -\frac{2}{3\eta} Q_{kl} + \lra{k}{l}
\ee
\be
-\frac{1}{N} \pardiff{W_{s}}{C_{kl}} = 0
\ee
\be
-\frac{1}{N} \frac{{\partial}^{2} \ln{{\cal L}}}{\partial {\eta}^{2}} = \frac{4}{{\eta}^{2}}+ 
\frac{2}{3 {\eta}^{2}} \vev{W|Q|W}
\ee
\be
-\frac{1}{N} \pardiff{W_{s}}{\eta} = -\frac{2}{3\eta} (QW)_{s}
\ee
\be
-\frac{1}{N} \pardiff{W_{l}}{W_{s}} = Q_{sl} \lab{matrix.last}
\ee

The inverse of the matrix $B$, given by equations\r{matrix.first} 
--\r{matrix.last}, describes the variances and covariances of 
the parameter estimates (see eq.\r{cov}).
Note that no observational errors have been included in this treatment
and that the uncertainties in the model parameters scale with the 
square root of the size
of the stellar sample. 
We now adopt the frame of reference for which the axes are aligned with the 
axes of the velocity ellipsoid i.e. where $C_{ij}$ is diagonal.
We then find that the errors in $\eta$, $W_{i}$, and $C_{ij}$ are given by
\bea
\makebox[1.5cm][l]{$\displaystyle \frac{var(\eta)}{{\eta}^{2}}$} & = & \frac{\al}{N} \lab{var.eta} \\
\makebox[1.5cm][l]{$\displaystyle \frac{var(W_{i})}{C_{ii}}$} & = & \frac{1}{N} \left( 1 + \frac{4}{9} Q_{ii} W_{i}^{2}\al \right) \lab{var.w} \\
\makebox[1.5cm][l]{$\displaystyle \frac{var(C_{jj})}{C_{jj}^{2}}$} & = & \frac{2}{N} \left( 1 + \frac{8}{9}\al \right) \lab{var.Cjj} \\
\makebox[1.5cm][l]{$\displaystyle \frac{var(C_{ij})}{C_{ii} C_{jj}}$} & = & \frac{1}{N} \lab{var.Cij}
\eea
where
\be
\al \equiv {\left( \frac{4}{3}+ \frac{2}{9} {\kappa}^2 \right) }^{-1} \;\;\;\;\;\; {\kappa}^2 \equiv \vev{W|Q|W}.\lab{alpha}
\ee
Here $\kappa$ is the generalization of the Mach number to an anisotropic
distribution.

We have derived fully analytic estimates of the errors in all parameters.
Note that the result given in equation\r{var.eta} can be written as
\be
\frac{\Delta \eta}{\eta} = N^{-\frac{1}{2}} {\left( \frac{4}{3}+\frac{2}{9} {\kappa}^2 \right)}^{-\frac{1}{2}}. \lab{var.eta.root}
\ee
That is, the error in the distance estimate (eqs.\ \r{var.eta.root} 
and\r{var.eta}) for the 
general case has the same form as the error for an isotropic distribution 
derived using a few basic assumptions (eq.\ \r{eta.error4}). 

\subsection{Non-Gaussian Velocities}

	As we show in \S 6, the actual distribution of RR Lyrae stars is
very far from Gaussian.  On the other hand, the formalism presented in
\S 4 explicitly assumes that the velocities are Gaussian.
That is, the likelihood given in equations\r{prob.fun1} and\r{prob.fun2} is 
the probability distribution for a 3-dimensional Gaussian.  
At first sight this appears to be a very serious problem because the true 
form of the distribution is only
crudely determined from the data and there is no a priori argument by which
one knows even how to parameterize the distribution.  We address this problem
in two ways.  First we argue in this section that one does not
introduce a bias by using a Gaussian likelihood function regardless of
the form of the actual velocity distribution.  (The use of a Gaussian function
{\it does} cause one to incorrectly estimate the uncertainties of the 
luminosity measurement, but by a calculable and -- as it turns out -- small
amount.)\ \ Second we confirm this result by Monte Carlo simulations in
\S 6.

	Why does the assumption of a Gaussian distribution not
introduce a bias? 
Early versions of the statistical parallax methods were explicitly based
on the first and second moments of the velocity distribution and as such were
intrinsically insensitive to the form of the stellar distribution. 
In the maximum likelihood method we should ideally
choose the velocity distribution function that matches the actual one.
For the case where there is not enough information about the 
underlying velocity distribution, it is desirable to pick a Gaussian.
This is because the Gaussian likelihood has the special characteristic that 
it works in essence by determining the means and dispersions of the velocities
(in each of three dimensions) separately from the radial velocity and 
proper-motion measurements and then forcing these to be equal by fixing the 
luminosity. The means and dispersions (and also covariances) are then reported
as nine additional parameters of the fit.  The method therefore effectively 
reproduces the naive procedure outlined in \S 2, which was based on the first 
and second moments of the distribution.
That is why it also reproduces the results derived using that procedure.  
It is straightforward to show that if one uses maximum likelihood to fit a 
non-Gaussian distribution to a Gaussian function 
parameterized by mean $W$ and variance $\sigma^2$, then the resulting values
of $W$ and $\sigma^2$ will be unbiased estimators of the mean and variance
of the (non-Gaussian) distribution\footnote{To be more precise $\sigma^2$
will be an unbiased estimator of $(N-1)/N \times$ the variance, but this
slight difference has no practical impact on the discussion here.}.  
Thus, adjusting the luminosity to maximize
the Gaussian likelihood of a non-Gaussian distribution still amounts to
equating the means and variances of the (non-Gaussian) distribution.  Since
the determinations of these means and variances are unbiased, so is the
estimate of the luminosity.

	If the underlying distribution is non-Gaussian, the Gaussian maximum likelihood
procedure will return estimates of the errors that {\it differ} from
the true errors.  Consider for example the simple isotropic model with 
dispersion $\sigma$, kurtosis $K$, bulk motion $W$, and negligible 
observational errors.  The maximum likelihood estimate of the error in $\eta$ is
(see eq.\r{var.eta})  $(\Delta \eta/\eta)^{-2} =(2/9)N[6 + {\kappa}^2]$ while
the true error is (see eq.\r{eta.error4}) 
$(\Delta \eta/\eta)^{-2} =(2/9)N[(12/[K-1]) + {\kappa}^2]$. 
For a more general velocity distribution that is the product of distributions
in the $\pi$, $\theta$, and $z$ directions with dispersions 
$({\sigma}_{\pi},\sigma_{\theta},\sigma_z)$ and kurtoses $(K_{\pi},K_{\theta},K_z)$ 
(and assuming that the bulk motion is in the $\theta$ direction), 
equation\r{var.eta} yields 
$(\Delta \eta/\eta)^{-2} =(2/9)N[6 + (W/\sigma_{\theta})^2]$. Generalizing from 
equation\r{eta.error4}, we estimate the true errors as 
$(\Delta \eta/\eta)^{-2} =(2/9)N[(4/(K_{\pi}-1)+4/(K_{\theta}-1)+4/(K_z-1) + 
(W/\sigma_{\theta})^2]$.
(We confirm this estimate numerically and mention some practical complications
in \S 6.)\ \ For an arbitrary
velocity distribution, the true and maximum likelihood-estimated errors could in principle
differ substantially.  However, for the actual RR Lyrae population,
$4/(K_{\pi}-1)+4/(K_{\theta}-1)+4/(K_z-1) \approx 7$, close to the Gaussian value 
of 6.  
This implies that the maximum likelihood estimated errors are nearly 
equal to the true errors.

\section{Complete analysis}

In this section we obtain the formulae needed to analyze real 
data and to carry out Monte Carlo simulations.
Our analysis here is similar to the one conducted in \S 4, but we
account for a few additional effects.
 
The probability density of finding a star with velocity components in the
ranges $(u_{i}, u_{i}+du_{i})$ is now given by:
\be
L(u_{i}; \eta, C_{ij}, w_{i})= \frac{{\eta}^{2}}{{(2\pi|M|)}^{\frac{1}{2}}} 
\exp{\left[-\frac{1}{2} {\sum}_{i,j} (s_{i}-w_{i}){(M^{-1})}_{ij}(s_{j}-w_{j})\right]}, \lab{prob.fun3}
\ee
where $w_{i}$ is the bulk motion (defined more precisely below), and
$s_{i}$ is the stellar velocity expressed in its {\it local} Galactic
coordinate frame (the velocity that would be measured by an observer located
at the star's position and at rest with respect to the Galactic center)
\be
s_{i}={\sum}_{j=1}^{3} R_{ij} \left( u_{j}-v_{\odot j} \right), \lab{z.def}
\ee
with ${\bf v_{\odot}}=(-9,232,7) \;\rm{km} \, {\rm s}^{-1}$.

The matrix $M$ is the full velocity covariance matrix, defined as
\be
M_{ij}=C_{ij} + {\sigma}_{r}^{2} r_{i} r_{j} + 
{\eta}^{2} {\sigma}_{t}^{2} P_{ij}. \lab{M.def}
\ee
Here ${\sigma}_{r}$ is the observational error in the radial velocity, 
${\sigma}_{t}$ is observational error in each of the inferred transverse
velocity components, and $P_{ij}$ is the projection operator on the plane of 
the sky. In equation\r{z.def} we explicitly subtract the velocity of the Sun 
${\bf v_{\odot}}$ from the star's velocity.
The velocity ${\bf v_{\odot}}$ is the sum of the velocity of the Sun's LSR 
${\bf v}_{\rm LSR}$ and the peculiar motion of the Sun relative to its LSR 
${\bf v_{\odot , {\rm LSR}}}$.
The completely new element not present in formula\r{prob.fun2} (nor
in any previous 
statistical parallax analyses of which we are aware) is the 
rotation
operator $R_{ij}$ that accounts for the fact that the star's velocity is drawn 
from a certain distribution (e.g., three-dimensional Gaussian) in the Galactic 
frame of reference centered at the star which is in general a very different 
distribution when expressed in terms of the Galactic frame centered at the Sun.
The matrix elements of the rotation operator are functions of the star's 
position and the distance from the Sun to Galactic center. We shall assume 
that the distance to Galactic center is $\sim 8 \, {\rm kpc}$.
The bulk motion, $w_{i}$, is relative to the {\em local} Galactic frame.
That is, in the Galactic frame at the Sun's position, the bulk motion is 
${\sum}_{j} {\left(R^{-1} \right)}_{ij} w_{j}$.
As in the no-error case, the logarithm of the total probability of finding the
stellar system in the observed state is the product of probability functions 
of the form\r{prob.fun3} for all the stars: 
$\ln{{\cal L}} = {\sum}_{k=1}^{N} \ln{L_{k}}$.

We generalize equation\r{Q.def1} to
\be
Q \equiv M^{-1}. \lab{Q.def2}
\ee
We maintain the decomposition\r{u.def1}, 
$u_{i} \equiv u_{r} r_{i} + \eta t_{i}$,  and we introduce the short notation
\be
y_{i} \equiv  {\sum}_{j} R_{ij}t_{j};\;\;\;\;\;\;\;\;\; x_{i} \equiv {\sum}_{j} Q_{ij}
s_{j}. \lab{three.def}
\ee

As previously, we use Dirac notation\r{Dirac} to designate scalar products 
and the symbol $\lra{k}{l}$ means ``add the terms with $k$ and $l$ exchanged, 
but only if $k \ne l$''.

The first derivatives of $\ln{L}$ are:
\bea
-\frac{\p \lnl}{\p \eta} & = & - \frac{2}{\eta} + \vev{y|x} - \eta{\sigma}_{t}^{2}\vev{x|P|x} + 
\eta {\sigma}_{t}^{2}{\rm tr}(Q P) \lab{dL.deta} \\
-\frac{\p \lnl}{\p W_i} & = & -x_i \\
-\frac{\p \lnl}{\p C_{k l}} & = & \frac{1}{2}(Q_{k l} - x_k x_l) 
 + \lra{k}{l} 
\eea

Second differentiation leads to:
\bea
\makebox[2cm][l]{$-\frac{\p^{2} \lnl}{\p {\displaystyle \eta}^2}$} & = & \frac{2}{{\eta}^2} + \vev{y|Q|y}
- {\sigma}_{t}^{2}\vev{x|P|x}-4\eta{\sigma}_{t}^{2}\vev{y|QP|x}+ 
{\sigma}_{t}^{2}{\rm tr}(Q P) \nonumber \\
& & - 2\eta^2{\sigma}_{t}^{4}{\rm tr}(Q P Q P) + 4\eta^2 
{\sigma}_{t}^{4}\vev{x|PQP|x} \\
\makebox[2cm][l]{$-\pardif{w_i}{w_j}$} & = & Q_{i j} \\
\makebox[2cm][l]{$-\pardif{C_{k l}}{C_{m n}}$} & = &
Q_{l m} x_n x_k - \frac{1}{2} Q_{l m}Q_{n k} + \lra{l}{k} + \lra{m}{n} \\
\makebox[2cm][l]{$-\pardif{\eta}{w_i}$} & = & -(Q y)_i + 
2\eta{\sigma}_{t}^{2}(Q P x)_i \\
\makebox[2cm][l]{$-\pardif{\eta}{C_{k l}}$} & = & -x_k(Q y)_l - 
\eta{\sigma}_{t}^{2}(Q P Q)_{k l} 
+2\eta {\sigma}_{t}^{2}(Q P x)_k x_l + \lra{l}{k}  \\
\makebox[2cm][l]{$-\pardif{w_i}{C_{k l}}$} & = & Q_{l i} x_k + \lra{l}{k} \lab{d2L.dwdC} 
\eea

In deriving these results we have made use of the identities
\be
\frac{\partial \ln{|M|}}{\partial \lambda} = {\rm tr} \left(M^{-1} \frac{\partial M}{\partial \lambda}\right),\;\;\;\;\;\;\;\;\;\;\;\; \frac{\partial M^{-1}}{\partial
\lambda}= - M^{-1} \frac{\partial M}{\partial \lambda} M^{-1}. \lab{iden}
\ee

We apply equations\r{dL.deta} --\r{d2L.dwdC} and Newton's method to find the 
maximum likelihood solution for the Layden et al. (1996) sample of halo RR 
Lyrae's. Table 1 gives the basic results and illustrates the process of 
correcting them for detected biases. 
Our ``assumed'' distance scale is the one derived by Layden et al. (1996).
As a result, if our determination were in perfect agreement with Layden et 
al. (1996), the scaling parameter $\eta$ should be exactly equal to 1.
The four rows give the maximum likelihood
values of the ten parameters characterizing the sample under four different
sets of assumptions.
None of the rows corresponds exactly to the results of Layden et al.
(1996), but the first row with $\eta$ increased by $\sim 0.015$ does so
approximately. The first row is not corrected for rotation or biases. 
The values presented in the second row are obtained using a non-unit rotation
operator (see discussion following eq.\r{M.def}).
Including rotation increases $\eta$ by about $0.4 \%$, only slightly influences
the bulk motion, but makes the velocity distribution less triaxial
($C_{22}^{\frac{1}{2}}$ closer to $C_{33}^{\frac{1}{2}}$) and more 
elongated (higher $C_{11}^{\frac{1}{2}}$).
The third row gives our best estimate for the values of the parameters
after correcting for the biases discussed in \S 6. We also take into account
that maximum likelihood underestimates the velocity variances by a factor 
$N/(N-1)$ as mentioned in \S 4.1.
Finally the fourth row takes into account the scatter in the absolute 
magnitudes of RR Lyrae stars.  The direct effects are estimated
numerically by introducing artificial scatter in the Monte Carlo simulations
and are in good agreement with the estimate given in \S 3.1.
The indirect effects of Malmquist bias are as described analytically
in \S 3.2.


The stars that enter our analysis are those defined as ``Halo-3'' population 
from Table 3 in Layden et al.\ (1996). Also, we use Layden's best estimate of 
the proper motion errors ($6.5 \, \rm mas\,yr^{-1}$) for stars taken from
the catalog compiled by Wan, Mao \& Ji (1980) rather than the value 
($5 \, \rm mas\,yr^{-1}$) artificially adopted by Layden et al. (1996) to match
their own sample (although this makes almost no difference).
We assume that RR Lyrae stars follow the absolute magnitude-metallicity relation 
($M_{V}= {\rm const} + 0.15 {\rm [Fe/H]}$) of Carney, Storm \& Jones (1992).
However, the results are only sensitive to the value of the absolute magnitude
at the mean metallicity of the sample, $\left< {\rm [Fe/H]} \right> = -1.61$, 
and not to the slope 
of the relation. We checked that the solutions for different slopes are 
statistically indistinguishable from one another.
By taking account of the metallicity, we restrict the possible scatter in 
RR Lyrae absolute magnitudes to the intrinsic scatter at fixed metallicity.
To make the corrections discussed in \S 3, we adopt for the Layden et al.
sample of field Galactic RR Lyrae stars ${\sigma}_M=0.15$.
To make this estimate, we first inspected color-magnitude diagrams of
several globular clusters and found that for these {\em very homogeneous}
populations the dispersion is typically ${\sigma}_M=0.08$. We take this
as a lower limit. The only available {\em inhomogeneous} sample at
approximately fixed distance is the 28 RRab Lyrae stars of Hazen \& Nemec 
(1992), which are 4$^{\circ}$ east of the LMC center and for which 
${\sigma}_M \sim 0.17$. We take this as an upper limit because some
of the scatter may be due to the dispersion in distance. For example,
if the LMC RR Lyrae population has an $r^{-3.5}$ profile and the flattening
$c/a=0.6$ (similar to the Galactic population), we find a dispersion
due to distance ${\sigma}_{M,dist}=0.09$, implying an intrinsic dispersion
of ${\sigma}_M \sim 0.14$.
In any event, those prefering other values of ${\sigma}_M$ should note that 
it is straightforward to find the corrected values of all parameters simply
by scaling the difference between rows 3 and 4.
The correction we apply in Table 1 is
\be
\frac{\delta \eta}{\eta} \approx 0.25 {\sigma}_M^2 - 0.64 {\sigma}_M^2 = -0.39 {\sigma}_M^2.
\ee
where the first term, estimated numerically, is due to the scatter in the 
absolute magnitude of RR Lyrae stars (\S 3.1) and the second term is due 
to Malmquist bias (\S 3.2).

To make connection to previous studies of Galactic structure, note that
it is customary to use the following notation:
\be
\sigma_U\equiv C_{11}^{\frac{1}{2}}, \;\;\;\;\;\;\;\;\;\;\sigma_V \equiv C_{22}^{\frac{1}{2}}, \;\;\;\;\;\;\;\;\;\;\sigma_W\equiv C_{33}^{\frac{1}{2}}.
\ee
Notice, however, that by using the non-unit rotational operator $R$ we are
actually measuring the underlying velocity distribution in the {\it local} 
Galactic frames of the stars in the sample under the assumption of 
Galactic axisymmetry. 
We therefore use $(\pi, \theta, z)$ rather than $(U,V,W)$.
The rotation operator $R$ is the appropriate first order correction to the
rectilinear solution (e.g., Layden et al. 1996) regardless of whether the
velocity distribution in the Galaxy is exactly axisymmetric.
Our best estimate of the RR Lyrae absolute magnitude at the mean metallicity 
of the sample $\left<\rm [Fe/H]\right> =-1.61$ is $M_V=0.75\pm 0.13$.
The velocity ellipsoid is $(\sigma_{\pi},\sigma_{\theta},\sigma_z)=(171\pm 10, 98\pm 8,
96\pm 8)\,\, \rm{km} \, {\rm s}^{-1}$ and the RR Lyrae population is moving
in $\theta$ direction at $-211 \pm 12 \,\, \rm{km} \, {\rm s}^{-1}$ relative
to the Sun, where the errors are adopted from YYTA case in the Table 3.

\section{Monte Carlo simulations}

Here we present the results of Monte Carlo simulations aimed at
checking and fine-tuning our analytic results.
There are two main classes of simulations. One class keeps all the star
positions from the Layden et al.\ (1996) sample unchanged giving results
closely related to those obtained using the real sample. For the second class,
the stars are placed randomly over the celestial sphere. The position-related 
biases can be determined by comparing the two classes of simulations.

In each simulation we construct 4000 mock samples of 162 halo stars
(162 is the size of the Layden et al. 1996 sample).
For each sample we generate a set of 162 space velocities drawn
from a distribution with specified means, dispersions and kurtoses
in the three principal directions.
For each star we transform its velocity components from the star's Galactic 
frame of reference to the Sun's frame and,
in some cases,  add Gaussian measurement errors in accordance with the
values given by Layden et al. (1996).

In the second step we find the most probable
parameters describing each of the samples. To analyze our mock
samples we use exactly the same maximum likelihood procedure that was used 
to obtain the results for the real stars. 


We perform several tests to check for the various systematic effects.
The results of these investigations are summarized in the Tables 2 and 3.
Table 2 gives the biases in $\eta$, $w_{i}$, and $C_{ii}^{\frac{1}{2}}$
found in each simulation.
The first column says whether the observational errors were included or were 
set equal to zero. The second column tells whether the velocities were 
rotationally adjusted (operator $R$ in equation\r{z.def}). The third 
column gives the information on whether the positions of the stars in the 
sample were the true ones from Layden et al. (1996) or were chosen randomly. 
The fourth column lists the assumed values of kurtoses in all three 
directions\footnote{Note that for YYTA case, the input kurtoses
are the ones that produce (in the mean) the same output values as those
obtained for the actual sample.}.
The fifth column assigns the name to each case.
The name contains the most important information about the case. For example
YYTH means that the observational errors were included in the analysis,
that velocities were rotationally adjusted, that the true positions of the
stars were considered, and that the kurtoses in all directions
were higher than Gaussian (here equal to 4). Generally, ``T'' stands for
true, ``R'' for random, ``G'' for Gaussian, ``H'' for high, ``L'' for low
and ``A'' for actual.
The next seven columns give the results for the most interesting parameters of 
the fit based on 4000 realizations. In all cases, the biases in the 
off-diagonal elements of the velocity covariance matrix (normalized in the 
same way as in the Table 1) are smaller than 0.01 and we therefore do not
display them here. The 
input values of the parameters for the underlying distribution are given 
below the descriptions of the column content.
The values in parentheses in the first row give the approximate errors
in the determinations of the biases based on the NNRG case.
The values in the table are the biases B detected for a given case
defined as: $B = ({\rm obtained \; value}) - ({\rm template \; value})$. 
The exceptions to this rule are the biases of the 
dispersion which are computed according to the scheme: 
$B = ({\rm obtained \; value}) -{[(N-1)/N]}^{\frac{1}{2}}  ({\rm template \; value})$. 
This comes from the fact that maximum likelihood returns variances that are 
$(N-1)/N$ times the values of the true ones (see \S 4.1). Note that with this 
definition of bias, one must subtract $B$ from the maximum likelihood solution
to get the corrected value.

The first five rows test the correctness of our implementation of the maximum
likelihood method. They contain the cases with Gaussian velocity
distributions ($K_{\pi}=K_{\theta}=K_{z}=3$) for which we expect the biases to be 
small. Indeed, the
deviations from the input values in the cases with the random positions
of the stars (NNRG and YYRG) are extremely small, that is, the obtained 
solutions are statistically indistinguishable from the input.
For the cases with the true positions of the stars (NNTG, YNTG and YYTG)
the deviations remain small but now they are statistically significant
and should be treated as biases rather than statistical fluctuations.

The next three rows, below the blank line, give the results for the
samples with velocities drawn from high- or low-kurtosis, non-Gaussian
distributions. One may suspect that our maximum likelihood procedure
derived assuming a  Gaussian velocity ellipsoid will not be very
accurate in such cases.
To the contrary, the determination of the parameters is extremely robust
and insensitive to the underlying velocity distribution.
The case YYTA is constructed to reproduce the input kurtoses of the
real sample of the Layden et al. (1996) stars. A glance at Figure 1 shows
that the distributions of the stars in the sample are highly non-Gaussian.
The measured kurtoses of RR Lyrae stars in three principal directions are:
$(2.18,3.14,3.93)$. The right panel of Figure 1 shows how
close the distribution in the radial direction is to a box distribution.
Analyzing our simulations we find that the most probable values of the
kurtoses of the true underlying distribution of RR Lyrae stars in the
solar neighborhood are $(K_{\pi}, K_{\theta}, K_{z})=(2.04,3.22,4.28)$.
Fortunately, as we see from the lower part of the Table 2, the parameter
determination is almost completely independent of the form of velocity 
distribution.


Table 3 contains the errors in the parameter determinations as obtained from 
the scatter of the Monte Carlo realizations.
For comparison the first row gives the errors predicted
analytically by equations\r{var.eta} --\r{var.Cij}. 
For the simulation with velocities with no measurement errors and unit matrix
operator $R$ (NNRG) our analytic estimates of the errors should agree with 
the scatter of the simulated results.
The agreement is striking and all the discrepancies are within statistical
uncertainties.
Additionally, accounting for the true positions of the stars (NNTG),
observational errors (YNTG) and non-unit rotation operator (YYTG)
affects the errors only slightly. Furthermore, the same is true
for the high-kurtosis (YYTH) or low-kurtosis (YYTL) cases.
The errors in the non-Gaussian cases cannot be predicted exactly from
the formulae given in \S 4.1 because the distribution is not
isotropic. Nevertheless, the predicted trend is clearly confirmed. 
We conclude that the analytic estimates of the uncertainties constitute 
excellent approximations to the realistic case. Both observational errors and 
higher kurtosis slightly increase the uncertainties.


In Table 4, we compare the kurtoses measured in the three principal directions
to those of the underlying velocity distributions.
First note that unless the underlying kurtosis is very low (as in the
NYRL case), the measured kurtosis is underestimated due to the finite size
of the sample (e.g., NNRG and NYRH cases). Additionally, measurement errors
tend to Gaussianize the distribution, and as a result reduce kurtoses
in high-kurtosis cases (e.g., YYRH vs. NYRH), increase kurtoses in low-kurtosis
cases (e.g., YYRL vs. NYRL), and have little effect on a Gaussian distribution
(e.g., YYRG vs. NNRG). The measured kurtoses are also biased due to the
non-isotropic positions of the stars in the sample on the sky (e.g.,
NNTG vs. NNRG or YYTH vs. YYRH), but this effect is relatively less
important for non-Gaussian distributions.
Our best estimate for the underlying kurtoses of the halo RR Lyrae
stars, based on the simulation results presented in Table 4, is 
$(2.04, 3.22, 4.28)$ (YYTA).


\section{Kinematics of Metal-Poor Halo Stars}

        As we emphasized in \S 2.1, the factor that fundamentally limits
the precision is the size of the sample.  This limitation is severe because
to expand the sample requires first identifying the new RR Lyrae stars and then
measuring their proper motions.  Some additional stars could undoubtedly be
found in the South with distances similar to those characteristic of the
sample.  In most cases, however, new proper motion studies would be required.
Moreover, the increase in the sample size would be modest.
To really increase the size of the sample substantially
would require finding stars at greater distances.   In the North, at least,
proper motions would be available from the Lick Proper Motion study.
However, the size of the errors for these stars would begin to approach those 
of the proper motions themselves, increasing the
possibility of systematic errors.  Thus, this route would appear to require
substantial additional work, lasting perhaps several decades.  In addition,
stars that are found too far from the Sun may have different kinematics which
would introduce additional systematic errors.

        There is, however, a poor man's route to increased statistics which
has the side-benefit of yielding new kinematic information about the stellar
halo.  If one repeats the analysis of \S 2, but with different numbers
of stars with radial velocity ($N_r$) and proper motion ($N_t$) measurements, 
one finds
\be
\biggl({\Delta \eta\over \eta}\biggr)^{2} =
{3\over 2}(6 + \kappa^2)^{-1}\biggl({2\over N_r} + {1\over N_t}\biggr)
\lab{combined1}
\ee
in place of equation\r{eta.error4}.
If one measured only the radial velocities for a substantial new population
of RR Lyrae stars, one could hope to drive down the error by up to a factor
$3^{1/2}$ without obtaining any new proper motions.  In
fact, it is not even necessary that they be RR Lyrae stars.  The only 
requirement is that the radial-velocity stars and the proper-motion
stars have the same kinematics.  This is trivially satisfied if all the
stars are RR Lyrae, but it can also be attained through careful selection
of other stars.

        For this purpose, we turn to the catalog of 1936 
non-kinematically selected metal-poor ([Fe/H]$<-0.6$) stars of 
Beers \& Sommer-Larsen (BSL, 1995).  We select stars {\it both} from the
BSL sample and Layden et al.\ sample with [Fe/H]$\leq -1.5$.  This should
insure that the great majority are from the stellar halo, not the
thick disk.  (If the fraction of the thick disk stars differs between 
the two samples, this will introduce systematic errors, as we quantify below.)
Next, we exclude BSL stars classified as ``variable'' to avoid
overlap with the RR Lyrae sample and also to avoid stars with possibly poor
radial-velocity determinations.  Finally, we eliminate stars with estimated
distances $>3$ kpc.  The precise distances are not important for the 
analysis, but we do wish to eliminate stars that are outside the volume
from which the RR Lyrae stars are drawn since they may partake of different
kinematics.  These criteria yield a BSL sub-sample of 724 stars and a
Layden et al.\ sub-sample of 106 RR Lyrae (of which 103 are from the sample
of 162 stars analyzed elsewhere in this paper).  The BSL stars are 
overwhelmingly turn-off, giant-branch, upper-main-sequence, and 
horizontal-branch stars (in that order) and so have masses 
(or progenitor masses) very similar to those of RR Lyrae stars.  
Hence, they should be drawn from very nearly the same underlying population,
regardless of the process by which the stellar halo formed.


        The new combined non-kinematic
sample consists of 830 stars total: 724 stars come from the BSL 
sample and 106 RR Lyrae stars from Layden et al.
In Table 5 we compare the parameter and error determinations for the
combined non-kinematic and Layden et al.\ samples.
The results are corrected for magnitude dispersion, Malmquist bias,
and other effects described in \S 5 and \S 6.
The Malmquist bias correction for the non-kinematic sample is the 
same as the one applied in Table 1 of \S 5.  The magnitude dispersion
correction was determined based on numerical simulations. 
Consequently, we compare final results: 
the best values and uncertainties for the Layden et al. (1996) sample
are repeated from the final rows of Tables 1 and 3, and the corresponding data 
are presented for the non-kinematic sample.  
The two results appear quite consistent, but since they are based partly on 
the same data, a careful statistical analysis is required.

        The first point to note is that Table 5 gives only the 
{\it statistical} error, i.e., the error that arises primarily from the 
finite size of the sample and secondarily from measurement errors.
For the non-kinematic sample, there is an additional error arising from
the possible differences in the fraction of thick-disk stars in the two
sub-samples.  Let these fractions be $f_1$ and $f_2$, respectively. Let
$f= (f_1+f_2)/2$, and let $\Delta = (f_1-f_2)/f$.  Since the proper motions
are determined entirely from the Layden et al.\ sub-sample and the radial
velocities are determined almost entirely from the BSL sample, the thick disk
stars will generate an additional error $\delta \eta/\eta\sim g f\Delta$
where $(1-g)\sim 0.5$ is the ratio of the typical thick-disk to typical halo
speeds relative to the Sun.  
We estimate the thick disk fraction $f$ from the fact that the 
non-kinematic sample has a rotation speed of $35\,\kms$ (see Table 5).  Since
halo stars have approximately zero rotation (BSL), this net rotation
must be produced by thick-disk contamination.  Since the rotation speed of
the thick disk is of order $170\,\kms$ (Casertano, Ratnatunga, \& Bahcall 
1990), the contamination level is $f\sim 35/170\sim 0.2$.
The mean distance from the plane is very similar for the two sub-samples,
$1210\,$pc for the Layden et al.\ stars and $926\,$pc for the BSL stars.
We therefore estimate $g\sim \pm 0.2$, and hence 
$\delta \eta /\eta\sim \pm 0.02$.  Since this error is uncorrelated with the
statistical error, we add the two in quadrature to obtain our final estimate
for the non-kinematic sample $\eta=0.953\pm 0.054$.

        Next, to either compare or combine the two samples, we must
evaluate the correlation coefficient of the two determinations,
$\gamma\equiv c_{12}(c_{11}c_{22})^{-1/2}$.  Here $c_{i j}$ is the covariance
matrix of the two measurements, with $c_{11}= (0.057)^2$ (from Table 5)
and $c_{22}=(0.054)^2$ (derived in the previous paragraph).
In the appendix, we find $\gamma = 0.46$.  This is very close to 0.5, the value
one would naively guess because the radial-velocity measurements are almost
completely independent while the proper motions are almost completely
dependent.

        The $1\,\sigma$ expected difference between the two measurements
is therefore $(c_{11} + c_{22} - 2 c_{12})^{1/2}= 0.059$.  That is, the
two measurements, 
$\eta_1=0.980\pm 0.057$ corresponding to $M_V=0.75\pm 0.13$ at 
[Fe/H]$=-1.61$,
and
$\eta_2=0.953\pm 0.054$ corresponding to $M_V=0.79\pm 0.12$ at 
[Fe/H]$=-1.79$
differ by less than $1\,\sigma$.  It is therefore
appropriate to combine them.  We apply standard linear theory
(e.g., Boutreux \& Gould 1996 eqs.\ 2.5 and
2.8) to find a best estimate and error
\be
\!\!\!\!\!\!\!\! \eta = \eta_1 - {(\eta_1 - \eta_2)(c_{11}-c_{12})\over 
c_{11}+c_{22}-2 c_{12}}=0.965 \;\;\;\;\;\;
\biggl({\Delta\eta\over \eta}\biggr) =  \biggl[c_{11} - {(c_{11}-c_{12})^2\over
c_{11}+c_{22}-2 c_{12}}\biggr]^{\frac{1}{2}}=0.047,
\ee
corresponding to $M_V=0.77\pm 0.011$ at [Fe/H]$=-1.71$.

        Finally, we note that Table 5 contains the first complete solution
of the velocity ellipsoid of halo stars as determined from a non-kinematically
selected sample.  One important (though not unexpected) result is that
the stellar halo is not moving relative to the LSR in either the radial or
vertical directions.  From Table 5, the estimates would appear to be 
$3\pm 8\,\kms$ 
away from the Galactic center and $0\pm 5\,\kms$ toward the north Galactic
pole.  In fact, each of these values (and errors) should be augmented by
a factor $(1-f)^{-1}\sim 1.25$ because the contaminating thick disk stars
are known to be bound to the disk and so have zero mean motion in both
directions.  Hence the best estimates for the halo bulk motion in the 
two directions are $4\pm 10\,\kms$ (radial) and $0\pm 6\,\kms$ (vertical).
The rotation of the halo ($35\pm 8\,\kms$) is larger than in the
solution for the 162 RR Lyrae alone, but this is to be expected because
that sample was selected in part kinematically, that is by eliminating 
stars with the most prograde orbits (Layden et al.\ 1996).  

\section{Discussion and Conclusions}
        We have investigated many potential sources of systematic error in
the RR Lyrae absolute magnitude calibration by statistical parallax using
a combination of analytic and Monte Carlo techniques.  We find that all
corrections to previous results are small and, in particular, that the highly
non-Gaussian RR Lyrae velocity distribution does not bias the determination
at all even though (or rather, because) the method explicitly assumes a 
Gaussian distribution.
We find that the mean RR Lyrae absolute magnitude is $M_V=0.75\pm 0.13$ at 
the mean metallicity of the sample $\left<\rm [Fe/H]\right>=-1.61$ compared 
to $M_V=0.71\pm 0.12$ obtained by Layden et al.\ (1996).  The largest source
of difference comes from including Malmquist bias which makes our estimate
0.03 mag fainter (for our adopted scatter ${\sigma}_M=0.15$).  Most
of the rest of the difference (0.01 mag) comes from the other corrections for
scatter in the absolute magnitude (0.03 mag brighter for Layden et al.\ 
versus 0.01 mag brighter for us).  There are several other smaller effects
that reduce the difference by 0.01 mag.

We also analyze a semi-independent non-kinematically selected sample of stars 
with metallicities
[Fe/H]$\leq -1.5$ taken from Layden et al.\ and BSL
and find similarly $M_V=0.79\pm 0.12$ at $\left<\rm [Fe/H]\right>=-1.79$.
Additionally, this analysis yields measurements of the radial bulk
motion ($4\pm 10 \,\kms$) and 
vertical bulk motion ($0\pm 6 \,\kms$) of the halo relative to the LSR.

        Our principal result is therefore that the RR Lyrae absolute
magnitude calibration by statistical parallax is extremely robust and that
the statistical error (0.13 mag) should be taken at face value.  

        If one assumes that type ab RR Lyrae stars in the LMC are
(apart from different mean metallicity) similar to those in the Layden
et al.\ (1996) sample, then the distance modulus to the LMC is
\be
\mu_{\rm LMC} = \langle V \rangle_{0,\rm LMC}
-\{0.75 + 0.15({\rm [Fe/H]} + 1.61)\}, \lab{lmcdist}
\ee
where $\langle V \rangle_{0,\rm LMC}$ is the dereddened mean apparent
magnitude of RRab's at the distance of the {\it center of mass} of the LMC.
Ideally, the way to determine this quantity is to measure $V$ for a large
sample of RRab's in an annulus around the LMC bar.  Stars well away from the 
bar should be little affected by internal LMC extinction and the mean 
foreground extinction is reasonably well understood. Moreover, by taking the 
average of annulus, one would assure that the mean distance of the sample is 
equal the center-of-mass distance, regardless of the geometry of the RR 
Lyrae distribution. Such photometry should soon be available from ongoing 
microlensing surveys.  However, the three RR Lyrae samples currently available
are both smaller and less ideal.  One is Walker's (1992) RR Lyrae sample
drawn from 6 LMC clusters (excluding one foreground cluster) for which
$$\langle V \rangle_{0,\rm LMC} = 18.98\pm 0.03\qquad
(\langle\rm [Fe/H]\rangle = -1.9)\qquad ({\rm clusters}).$$ 
The error is from the scatter and hence
accounts for measurement errors and uncertainty in the mean distance of the
clusters relative to the center of mass.  We estimate an additional error
in the mean extinction of 0.03.  Applying equation\r{lmcdist}, we find
$\mu_{\rm LMC} = 18.28\pm 0.14$.  However, two lines of evidence suggest
that cluster RR Lyrae stars may be systematically brighter than those in
the field.  First, Sweigart (1997) has argued that abundance
anomalies seen in cluster giants but not in halo field giants could be due
to rotational mixing in the former.  Such mixing would dredge up helium in
horizontal branch stars, making them up to a few tenths of a magnitude 
brighter.  Second, new main-sequence-fitting distances to Galactic globulars
(Reid 1997; Gratton et al.\ 1997) would, if confirmed, imply that cluster RR 
Lyrae stars are several tenths of a magnitude brighter than the value for
field stars reported here.  Hence, it seems prudent to restrict the comparison
to LMC {\it field} RR Lyrae stars.

        Two photometric studies of LMC field RR Lyrae stars have been 
conducted, both in $B$.  Graham (1977) found $\langle B \rangle=19.61\pm 0.02$
for 60 field RRab stars near NGC 1783, about $4^\circ$ northwest of the bar.
The error reflects only the dispersion in the measurements and Graham (1977)
notes that calibration errors are possible.  There is no consensus on the
reddening toward this field.  We adopt the estimate of Alvarado et al.\ 
(1995), $E(B-V)=0.08\pm 0.02$.  From the period-amplitude diagram of these
stars, Hazen \& Nemec (1992) estimate a mean metallicity 
$\langle{\rm [Fe/H]}\rangle=-1.3$.  To estimate the mean color, we use
the relation $(B-V)_0 = 0.658  + 0.710\log P + 0.097$[Fe/H] from 
Caputo \& De Santis (1992) which, from their Figure 5b, has an uncertainty
of $\pm 0.015$ for each individual star.  The mean period of the Graham 
(1977) stars is $\langle P\rangle=0.564\,$ days, implying $(B-V)_0=0.36$.
Using $R_V\equiv {\rm A_V/E(B-V)} = 3.1$, we then find
$$\langle V\rangle_{0,\rm N1763\,field} = 18.92 \pm 0.08,$$
where no account has been made for a possible calibration error.

        Hazen \& Nemec (1992) found $\langle B \rangle=19.61\pm 0.03$
for 28 field RRab stars near NGC 2210, about $4^\circ$ east of the bar.
They believe that their photometry is well calibrated.  They adopt 
$E(B-V)=0.08\pm 0.01$ from Caldwell \& Coulson (1985), and find 
$\langle{\rm [Fe/H]}\rangle=-1.8$ from the period-amplitude diagram.  
The mean period is $\langle P\rangle=0.576\,$days, which implies
$(B-V)_0 = 0.31$.  Hence  
$$\langle V\rangle_{0,\rm N2210\,field} = 18.98 \pm 0.05,$$

        Combining the apparent magnitude and metallicity measurements with
the absolute magnitude of RR Lyrae stars yields distance-modulus estimates
$\mu_{\rm N1763\,field}=18.18\pm 0.15$ and $\mu_{\rm N2210\,field}=18.26\pm 14$
which seem quite consistent.  However, unless the distribution of RR Lyrae
is spherical, these fields will not be at the same distance as the LMC
center of mass.  Assuming that (like the Galactic RR Lyrae stars) they have
a radial profile $\propto r^{-3.5}$ and an axis ratio $c/a=0.6$, and 
taking the inclination of the LMC to be $i=27^\circ$ (Bessel, Freeman 
\& Wood 1986), we find offsets of $+0.05\,$mag and $-0.05\,$mag for the two 
fields, respectively.  This leads to two estimates for the LMC center of mass
$\mu^{\rm N1763\,field}_{\rm LMC}=18.13\pm 0.15$ and
$\mu^{\rm N2210\,field}_{\rm LMC}=18.31\pm 0.14$.  
Since the errors of these
determinations are correlated, the difference is $\Delta \mu = 0.18\pm 0.09$,
which is uncomfortably large.  While both of these determinations are
much smaller than the traditional value $\mu_{\rm LMC}=18.5$ and are smaller
yet compared to several new determinations (Reid 1997; Gratton et al.\ 1997;
Feast \& Catchpole 1997), there remains the possibility that the underlying 
measurements of $\langle V\rangle_0$ are affected by systematic errors.
A robust estimate of this quantity should therefore await results from the 
microlensing surveys.

\clearpage

\appendix

\setcounter{equation}{0}
\renewcommand{\theequation}{\thesection\arabic{equation}}
\section{Determination of the Correlation Coefficient}

We now return to the framework of \S 2 in order to evaluate the correlation
coefficient $\gamma\equiv c_{12}(c_{11} c_{22})^{-1/2}$.  The error for
a single determination of $\eta$ has the form given by equation\r{combined1}. 
For two partially dependent determinations of $\eta$, this generalizes to
\be
c_{i j} = {3\over 2}(6+\kappa^2)^{-1}(2 c^r_{i j} + c^t_{i j})
\lab{appendix1}
\ee
where
$c_{i j}$ is the $2\times 2$ covariance matrix of the errors in $\eta$,
\be
c^r = \left( \begin{array}{cc} 
\frac{1}{N_{r,1}} & \frac{N_{r,12}}{N_{r,1}N_{r,2}} \\
\frac{N_{r,12}}{N_{r,1}N_{r,2}} & \frac{1}{N_{r,2}}
\end{array} \right),
\ee
and similarly for $c^t$.  Here $N_{r,1}= 162$ is the number of radial-velocity
stars in the Layden et al. sample, $N_{r,2}=830$ is the number in the 
non-kinematic sample, and $N_{r,12}=103$ is the number of overlap stars.  
Similarly, $N_{t,1}=162$, $N_{t,2}=106$, and $N_{t,12}=103$.  It is 
immediately clear that the diagonal elements of equation\r{appendix1} reduce 
to equation\r{combined1}. The off-diagonal terms take account of the 
covariance between the two radial-velocity-based velocity ellipsoids and the 
covariance between the two tangential-velocity-based velocity ellipsoids.  
However, equation\r{appendix1} takes account only of the statistical errors.  
We now take account of the additional systematic error caused by differering
levels of thick disk contamination.  Recall that this error affects only
the non-kinematic determination, so it is included by adding a matrix to 
equation\r{appendix1}
\be
c_{i j} = {3\over 2}(6+\kappa^2)^{-1}(2 c^r_{i j} + c^t_{i j}) +
(g f \Delta)^2\left( \begin{array}{cc} 
0 & 0 \\
0 & 1
\end{array} \right),
\ee
where $g f \Delta=0.02$ was evaluated in \S 7.
Hence, $\gamma=c_{12}(c_{11}c_{22})^{-1/2}=0.46$.

\clearpage

\figcaption[fig1.ps]
{Left panel shows the velocity distribution functions for three Galactic 
directions. Right panel shows that the distribution in radial direction is 
very boxy. \label{fig1}}

\clearpage

\begin{table}
\dummytable\label{table1}
\end{table}

\begin{table}
\dummytable\label{table2}
\end{table}

\setcounter{table}{4}
\begin{table}
\dummytable\label{table5}
\end{table}
\setcounter{table}{2}

\begin{deluxetable}{ccccccccccc}
\tablewidth{0pt}
\tablecaption{Errors for the cases considered in our Monte Carlo simulations
\label{table3}}
\tablehead{
\colhead{Case name} & \colhead{$\Delta \eta$} & \colhead{$\Delta w_{1}$} & \colhead{$\Delta w_{2}$} & \colhead{$\Delta w_{3}$} & \colhead{$\Delta C_{11}^{\frac{1}{2}}$} & \colhead{$\Delta C_{22}^{\frac{1}{2}}$} & \colhead{$\Delta C_{33}^{\frac{1}{2}}$} & \colhead{$\Delta {\tilde{C}}_{12}$} & \colhead{$\Delta {\tilde{C}}_{13}$} & \colhead{$\Delta {\tilde{C}}_{23}$}
}
\startdata
     & 0.052 & 13.20 & 10.67 & 7.55 & 10.98 & 6.53 & 6.27 & 0.079 & 0.079 & 0.079 \nl
&&&&&&&&&&\nl
NNRG & 0.052 & 12.95 & 10.64 & 7.61 & 10.99 & 6.47 & 6.19 & 0.079 & 0.079 & 0.080 \nl
NNTG & 0.053 & 12.87 & 11.26 & 7.63 & 11.62 & 6.69 & 5.79 & 0.078 & 0.080 & 0.080 \nl
YNTG & 0.057 & 13.26 & 11.84 & 8.08 & 12.14 & 7.24 & 6.29 & 0.086 & 0.086 & 0.091 \nl
YYTG & 0.056 & 13.25 & 11.73 & 8.08 & 12.09 & 7.21 & 6.28 & 0.085 & 0.086 & 0.090 \nl
YYRG & 0.054 & 13.31 & 10.97 & 8.16 & 11.45 & 7.01 & 6.81 & 0.085 & 0.087 & 0.091 \nl
&&&&&&&&&&\nl
YYTH & 0.059 & 13.28 & 11.96 & 8.12 & 13.98 & 8.26 & 7.39 & 0.086 & 0.086 & 0.091 \nl
YYTL & 0.055 & 13.60 & 11.62 & 8.12 & 10.20 & 6.20 & 5.12 & 0.088 & 0.086 & 0.089 \nl
YYTA & 0.057 & 13.59 & 11.83 & 8.14 & 10.37 & 7.50 & 7.64 & 0.086 & 0.084 & 0.090 \nl
\enddata
\tablecomments{As previously, the name of the case uniquely characterizes
statistical properties of velocity input. The first row, just below column 
description,  gives the errors predicted analytically in \S 4.}
\end{deluxetable}

\clearpage

\begin{deluxetable}{ccc}
\tablewidth{0pt}
\tablecaption{Comparison of measured kurtoses to those of underlying
velocity distribution. The effects of the finite size of the sample,
observational errors and non-isotropic positions of the stars in the sample
are presented \label{table4}}
\tablehead{
\colhead{Case name} & \colhead{Underlying kurtoses} & \colhead{Measured kurtoses}
}
\startdata
NNRG & (3,3,3) & (2.98, 2.99, 2.98) \nl
YYRG & (3,3,3) & (2.99, 3.05, 3.05) \nl 
NNTG & (3,3,3) & (2.99, 2.97, 3.01) \nl
YYTG & (3,3,3) & (3.01, 3.02, 3.05) \nl
&&\\
NYRL & (2,2,2) & (2.02, 2.03, 2.03) \nl 
YYRL & (2,2,2) & (2.14, 2.34, 2.36) \nl
YYTL & (2,2,2) & (2.16, 2.36, 2.31) \nl 
NYRH & (4,4,4) & (3.91, 3.91, 3.91) \nl 
YYRH & (4,4,4) & (3.82, 3.73, 3.72) \nl 
YYTH & (4,4,4) & (3.83, 3.66, 3.77) \nl 
YYTA & (2.04,3.22,4.28) & (2.18, 3.14, 3.93) \nl
\enddata
\end{deluxetable}

\clearpage

\end{document}